\documentclass[aps,prl,twocolumn,superscriptaddress,showpacs]{revtex4}
\usepackage{amssymb}

\usepackage{graphicx}

\draft

\begin{document}

\title{Anti-Resonance and the \textquotedblleft 0.7
Anomaly\textquotedblright\ in Conductance through a Quantum Point Contact }
\author{Ye Xiong}
\affiliation{Department of Physics, Oklahoma State University, Stillwater, Oklahoma 74078 }
\author{X. C. Xie}
\affiliation{Department of Physics, Oklahoma State University, Stillwater, Oklahoma 74078 }
\affiliation{International Center for Quantum Structures and Institute of Physics, The
Chinese Academy of Sciences, Beijing 100080, P.R. China}
\author{Shi-Jie Xiong}
\affiliation{Department of Physics, Nanjing University, Nanjing 210093, China}
\date{\today}

\begin{abstract}
We investigate the transmission of electrons through a quantum point contact
by using a quasi-one-dimensional model with a local bound state below the
band bottom. While the complete transmission in lower channels gives rise to
plateaus of conductance at multiples of $2e^{2}/h$, the electrons in the
lowest channel are scattered by the local bound state when it is singly
occupied. This scattering produces a wide zero-transmittance
(anti-resonance) for a singlet formed by tunneling and local electrons, and
has no effect on triplets, leading to an exact $0.75(2e^{2}/h)$ shoulder
prior to the first $2e^{2}/h$ plateau. Formation of a Kondo singlet from
electrons in the Fermi sea screens the local moment and reduces the effects
of anti-resonance, complementing the shoulder from 0.75 to 1 at low
temperatures.
\end{abstract}

\maketitle


The quantization of conductance in mesoscopic systems has been observed in
quantum point contacts (QPC) where a series of plateaus at multiples of $%
2e^{2}/h$ appears in curves of conductance $G$ versus gate voltage $V_{g}$ 
\cite{1,2}. Prior to the first integer plateau, a shoulder of 0.7($2e^{2}/h$%
) has often been observed and is called the \textquotedblleft 0.7 anomaly" 
\cite{3,4,5,6}. Considerable experimental and theoretical efforts have been
devoted to understanding this anomaly \cite{7,8,9,a1,a2}. In Ref. %
\onlinecite{a1} it is suggested that the anomaly originates from the 3:1
triplet-singlet statistical weight ratio if two electrons form bound states
by some attractive interaction and the triplet has a lower energy. From
calculations for two electrons in the Hubbard model or the Anderson
Hamiltonian, two resonance peaks in transmission spectrum are shown,
corresponding to singlet and triplet states with weights 0.25 and 0.75 \cite%
{a2}. Some attention has been focused on the Kondo effect in such systems 
\cite{10,11}. In a recent experiment the temperature and magnetic-field
dependence of the differential conductance was investigated in detail to
highlight the connection to the Kondo problem\cite{12}, and subsequently a
theoretical study was carried out in further support of Kondo physics in QPC 
\cite{13}.

Since relevant states undergo empty, single and double occupations even
within the width of one plateau or shoulder, including global charge
variations and fluctuations in calculations is important. The "0.7 anomaly"
appears only prior to the first integer plateau, and only at relatively
higher temperatures, for which the intensity of the Kondo effect is
negligible. The shape of the shoulder is much different from the two
resonance peaks of the singlet and triplet states \cite{a2}, and there is no
evidence of the attractive interaction from which two-electron bound states
can be formed \cite{a1}. To address the above mentioned questions, in this
Letter we theoretically study the transport through a quantum point
contact(QPC) by using a model which includes the Coulomb interaction, the
charge fluctuations, and the multi-channel structure on an equal footing. By
using the singlet-triplet representation to label spin states of the
tunneling electron and the local electron, we show a wide anti-resonance for
the singlet channel near the band bottom, giving rise to the 0.75 shoulder.
The shoulder is complemented to $1$ by the formation of a Kondo singlet at
low temperatures, and manifests itself at higher temperatures when the Kondo
singlet collapses. Thus, the role of the Kondo singlet is to suppress the
anti-resonance, quite different from the Kondo physics discussed in Ref. 
\cite{13}. A simple scaling curve for conductance is obtained and is found
to compare well with the experimental one. The results provide consistent
explanations for a wide range of characteristics observed in experiments.

A QPC can be described with a narrow and short bar connected to the left and
right leads which serve as reservoirs. Thus, one obtains several continuous
1D subbands. Some local levels, may be virtual bound states\cite{13}, can be
created by the specific QPC geometry. These states are isolated from the
leads and should not be included in the band continuum. The potential of the
bar area, including both the band continuum and the local levels, is tuned
by the gate voltage. The single-electron energies of subbands and levels are
shown in Fig. 1(a). We include on-level Coulomb repulsion $U$ of electrons
confined in one local state. In equilibrium the Fermi energy of reservoirs
is fixed and we set it as the energy zero. The occupations of levels is
controlled by tuning the gate voltage. In Figs. 1(b), 1(c) and 1(d) we show
the empty, single and double occupations of the level nearest the band
bottom. When the Fermi level crosses the band bottom, the first subband
contributes to the conductance and the first plateau appears. At this moment
most of the local levels are fully occupied and have no effect on the
transport, except the one closest to the band bottom which may be singly
occupied due to the on-level interaction. It is sufficient to include this
level in our model. By tuning $V_{g}$ further this level is also fully
occupied and the plateau structure is determined only by the subbands.

\begin{figure}[tbp]
\includegraphics[width=6.8cm]{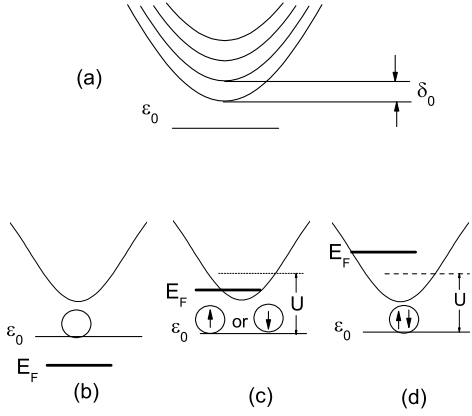}
\caption[FIG]{(a) Sketchy illustration of dispersion relation of 1D subbands
and local levels. $\protect\delta_0$ is energy spacing between subbands, and 
$\protect\epsilon_0$ is the position of the local level. (b) Empty level. $%
E_F$ denotes the Fermi level. (c) Singly occupied level. (d) Doubly occupied
level. }
\label{FIG.1}
\end{figure}

The Hamiltonian of the system can be written as 
\[
H=\sum_{m,i,\sigma }t_{0}(a_{m,i,\sigma }^{\dag }a_{m,i+1,\sigma }+\text{H.c.%
})
\]%
\[
+\sum_{m,i,\sigma }[(m-1)\delta _{0}-eV_{g}]a_{m,i,\sigma }^{\dag
}a_{m,i,\sigma }
\]%
\[
+\sum_{\sigma }t^{\prime }(a_{1,0,\sigma }^{\dag }d_{\sigma }+\text{H.c.}%
)+\sum_{\sigma }(\epsilon _{0}-eV_{g})d_{\sigma }^{\dag }d_{\sigma }
\]%
\begin{equation}
+Ud_{\uparrow }^{\dag }d_{\uparrow }d_{\downarrow }^{\dag }d_{\downarrow },
\end{equation}%
where $a_{m,i,\sigma }$ and $d_{\sigma }$ are the annihilation operators of
electrons on the $i$th site of the $m$th continuum channel and at the local
state, respectively, with $\sigma $ being the spin index. $\delta _{0}$ is
the energy spacing between subband bottoms. $\epsilon _{0}$ stands for the
position of the local level. $t_{0}$ is the hopping integral of the
channels, and $t^{\prime }$ is the coupling of the lowest channel and the
bound state at a site ($i=0$). Only the coupling of the local level to the
lowest channel ($m=1$) is considered, because local level coupling has no
effect on the transmission in higher channels due to the double occupation.

For $V_{g}\ll 0$ and at the zero temperature, the potential is too high and
the states in the bar, including the local state and the continuum chains,
are empty. With increasing $V_{g}$, the local state becomes singly occupied
in the range of $\epsilon _{0}<eV_{g}<\epsilon _{0}+U$. For $eV_{g}>\epsilon
_{0}+U$, it is doubly occupied and has no effect on the conductance. On the
other hand, the $m$th continuum channel gives $2e^{2}/h$ contribution to the
conductance if $eV_{g}>(m-1)\delta _{0}-2t_{0}$, because in this case the
Fermi level is higher than the bottom of this subband. This provides the
integer plateaus in the $G$-$V_{g}$ curves with plateau width $\delta _{0}/e$%
.

Since the local level is below the bottom of the whole band continuum, the
tunneling in the first channel is affected by its charge and spin states.
There are 4 states of the local level: the empty, $|\phi _{1}\rangle
=|0\rangle $, the spin up and down single occupation, $|\phi _{2}\rangle
=|\uparrow \rangle $ and $|\phi _{3}\rangle =|\downarrow \rangle $, and the
double occupation, $|\phi _{4}\rangle =|\uparrow \downarrow \rangle $. If an
electron is injected into the channel, it will be scattered by the state on
the local level. At first we consider only the tunneling and local electrons
and ignore the other electrons in the Fermi sea. These considered electrons
form a many-body state 
\begin{equation}
|\psi \rangle =\sum_{n=1}^{4}\sum_{i,\sigma }p_{n;1,i,\sigma }|\phi
_{n}\rangle \bigotimes |1,i,\sigma \rangle ,
\end{equation}%
where $|1,i,\sigma \rangle $ is the orbital at site $i$ with spin $\sigma $
in the first channel, and $p_{n;1,i,\sigma }$ is the corresponding
coefficient. By applying the Hamiltonian on $|\psi \rangle $ one obtains the
Schr\"{o}dinger equations for the coefficients. These equations can be
expressed with an equivalent single-particle network in which every site
represents a combination of indices $(n;1,i,\sigma )$ of the coefficients 
\cite{14}. In the present case the network is an 8-channel one where every
channel stands for a combination of $n$ and $\sigma $, and a site in a
channel corresponds to a coordinate $i$ in the chain. In the network the
channels with $(n=2,\sigma =\uparrow )$ and $(n=3,\sigma =\downarrow )$, and
the channels with $n=1,4$, corresponding to the empty and doubly occupied
local states, are independent. The other 2 channels are connected at site $%
i=0$ but can be easily decoupled with the transformation $|S,i\rangle =\frac{%
1}{\sqrt{2}}(|\phi _{2}\rangle \bigotimes |1,i,\downarrow \rangle -|\phi
_{3}\rangle \bigotimes |1,i,\uparrow \rangle )$ and $|T,i\rangle =\frac{1}{%
\sqrt{2}}(|\phi _{2}\rangle \bigotimes |1,i,\downarrow \rangle +|\phi
_{3}\rangle \bigotimes |1,i,\uparrow \rangle )$. The final network with 8
independent channels is shown in Fig. 2. For single occupation there is one
singlet channel with one scatterer (6) and three pure triplet channels (3,
4, and 5) without scattering by the local level. Different from Ref. \cite%
{a1}, the singlet-triplet notation used here is merely to label spin states
of the tunneling and local electrons that are not bound together. For $m>1$
and for the non-scattering pure channels of $m=1$, the transmission
coefficient is 
\begin{equation}
\tau _{m}(\epsilon )=\left\{ 
\begin{array}{l}
1,\text{ for }|\epsilon -(m-1)\delta _{0}+eV_{g}|<2t_{0}, \\ 
0\text{ otherwise. }%
\end{array}%
\right. ,  \label{trans}
\end{equation}%
where $\epsilon $ is energy of the injected electron. For the singlet
(empty) channel of $m=1$, 
\begin{equation}
\tau _{\text{s(e)}}=\frac{4[u_{\text{s(e)}}-2\cos (k)]^{2}\sin ^{2}(k)}{v_{%
\text{s(e)}}^{4}+4[u_{\text{s(e)}}-2\cos (k)]^{2}\sin ^{2}(k)}  \label{tran}
\end{equation}%
for $|\epsilon +eV_{g}|<2t_{0}$ and $\tau _{\text{s(e)}}=0$ otherwise, where 
$k$ is the momentum of the tunneling electron determined by $\epsilon
=2t_{0}\cos (k)-eV_{g}$, and $u_{\text{s}}=(\epsilon _{0}+U)/t_{0}$, $u_{%
\text{e}}=\epsilon _{0}/t_{0}$, $v_{\text{s}}=\sqrt{2}t^{\prime }/t_{0}$, $%
v_{\text{e}}=t^{\prime }/t_{0}$.

\begin{figure}[tbp]
\includegraphics[width=6.8cm]{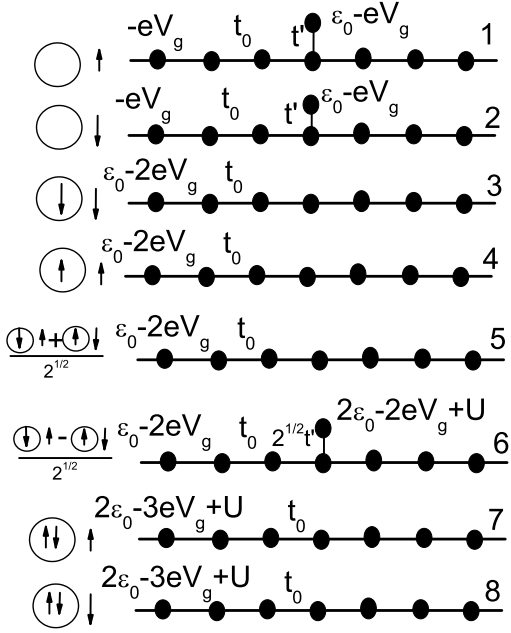}
\caption[FIG]{Eight independent channels for a tunneling electron. The state
of the local level is illustrated by circles, and the spin of the tunneling
electron is represented by an arrow outside the circle. }
\label{FIG.2}
\end{figure}

For the singlet channel the scatterer produces an anti-resonance at $%
\epsilon =\epsilon _{0}-eV_{g}+U$ with a zero transmission coefficient at
the minimum and with semi-width ${t^{\prime }{}^{2}}/t_{0}|\sin (k)|$. Near
the subband bottom where we focus, $\sin (k)\sim 0$, so the width of the dip
is extremely large even though $t^{\prime }$ may be much smaller than $t_{0}$%
. As a result, the dip of $\tau _{\text{s}}$  develops to a flat plateau
with nearly zero height. This is different from the results of Ref. \cite{a2}
where the singlet and triplet states give two separate resonance peaks with
different weights. Without the magnetic field and ignoring the effects of
other electrons, in the range of single occupation comma the conductance is
governed by $e^{2}/2h$ times the sum of the transmission coefficients of the
singlet and triplet channels which are non-zero only for the tunneling
electron with energy higher than the subband bottom. Here the prefactor $1/2$
stands for the weight of one spin state of the local electron. As mentioned
above, $\epsilon _{0}\lesssim -2t_{0}$, so $\epsilon _{0}+U$ may be above
the subband bottom. Thus, in the range of $-2t_{0}<eV_{g}<\epsilon _{0}+U$
conductance plateaus at height $0.75(2e^{2}/h)$. For $eV_{g}>\epsilon _{0}+U$%
, the conductance is dominated by the double-occupation channels (7 and 8 in
Fig. 2), leading to a jump from 0.75 to 1 at $eV_{g}=\epsilon _{0}+U$. This
is the origin of the 0.7 anomaly within our model.

\begin{figure}[tbp]
\includegraphics[width=8.5cm, height=7cm]{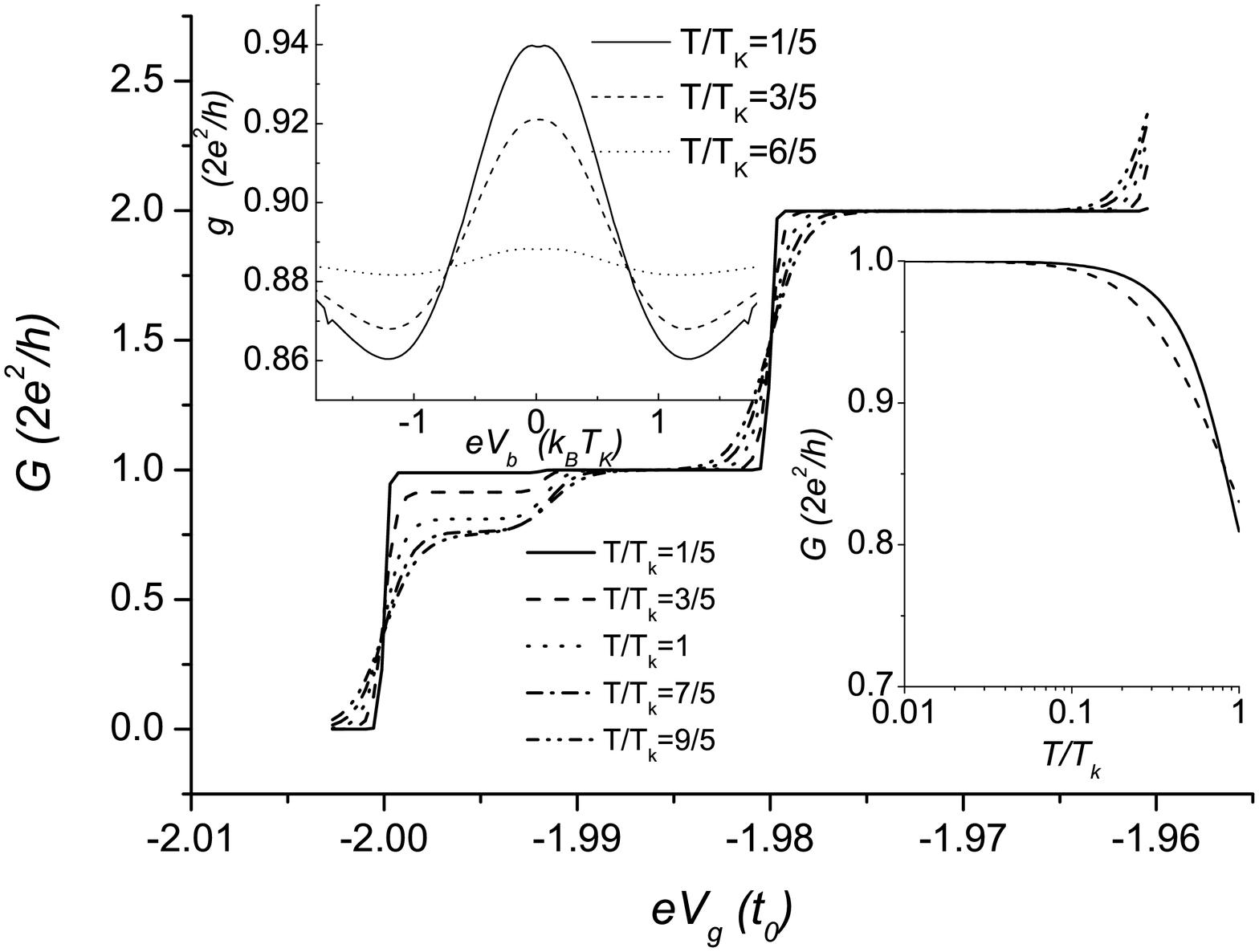}
\caption[FIG]{Linear conductance as a function of the gate voltage. The
parameters are: $\protect\delta_0=0.02t_0$, $U=0.016t_0$, $\protect\epsilon%
_0=-2.008t_0$, $T_K = 0.0005t_0 $ and $t^{\prime}=0.05t_0$. Upper inset:
Differential conductance as a function of bias voltage for $eV_g = -1.993t_0$%
. Lower inset: Linear conductance as a function of scaled temperature $T/T_K$%
. Solid line: The calculated conductance. Dashed line: The experimentally
fitted conductance.}
\label{FIG.3}
\end{figure}

Now we consider the effect of the electrons in the Fermi sea. It is known
that at low temperatures the electrons from Fermi sea are coupled with the
local electron to form a Kondo singlet. In this case the local
state is screened by the Fermi electrons and is no longer available for the
injected electron in the singlet channel, resulting in the disappearance of
the side-coupled scatterer for this channel. This removes the anti-resonance
and enhances the transmission coefficient from nearly zero to one. If the
probability of annihilating the Kondo singlet is $p_k$, the average
transmission coefficient of the singlet channel is $\bar{\tau_{\text{s}}} =
(1-p_k) \tau_1 +p_k \tau_{\text{s}} $. We choose $p_k=tanh(-T/T_{K})^{2} $.
This simple form reflects the physics that at low temperature the Kondo
singlet peak in the spectral density goes as $1-\alpha T^2$ \cite{Costi}.
The formation of the Kondo singlet has no effect for the triplet channels
since they are without scattering.

Combining all the contributions, the linear conductance at temperature $T$
can be calculated as 
\begin{equation}  \label{cond}
G(T) = -\frac{2e^2}{h} \int d \epsilon \frac{\partial f (\epsilon, T) }{%
\partial \epsilon} \tau(\epsilon),
\end{equation}
\[
\tau (\epsilon) = P_0(T) \tau_{\text{e}}(\epsilon) +P_2(T) \tau_1(\epsilon) 
\]
\[
+\frac{P_1(T)}{4} [\bar{\tau}_{\text{s}}(\epsilon)+3\tau_1(\epsilon)]+
\sum_{ m \geq 2 } \tau_m(\epsilon) ,  
\]
where $f(\epsilon,T)$ is the Fermi-Dirac distribution, $P_l(T)$ is the
thermal probability of finding the local level occupied by $l$ electrons,  
\[
P_{1}(T) = \frac{2}{Z} e ^{-\frac{\epsilon_0-eV_g}{k_B T} },\, P_{2}(T) = 
\frac{1}{Z}  e^{-\frac{2\epsilon_0-2eV_g+U}{k_BT}},  
\]
\begin{equation}
P_{0}(T)= \frac{1}{Z}, \, Z= 1 +2 e^{-\frac{\epsilon_0-eV_g}{k_BT} } + e^{ -%
\frac{2\epsilon_0-2eV_g+U}{k_BT}} .
\end{equation}

In Fig. 3 we plot the conductance as a function of the gate voltage for
several values of $T$. The local level $\epsilon _{0}$ is below and close to
the band bottom, $\epsilon _{0}+U$ is above it, and $t^{\prime }\ll t_{0}$,
reflecting the localized nature of the level. The value of $t_{0}/\delta _{0}
$ corresponds to the number of channels. A $0.75$ shoulder is clearly seen
at high temperatures $T>T_{K}$, meanwhile it is complemented to 1 at low
temperatures $T<T_{K}$. This accounts for the basic features observed in the
recent experiment \cite{12}.

At low temperatures $P_{l}$ and $f(\epsilon )$ can be approximated with step
functions and for $-2t_{0}<eV_{g}<\epsilon _{0}+U$, Eq. (\ref{cond}) becomes 
\begin{equation}
G(T)=\frac{2e^{2}}{h}\left( \frac{3}{4}+\frac{1}{4}F(\widetilde{T})\right) ,
\label{scal}
\end{equation}%
where $F(\widetilde{T})$ is a universal function of rescaled temperature $%
\widetilde{T}=T/T_{K}$, describing the influence of the Kondo effect. Here,
the range of $F$ is $[0,1]$, the prefactor of $F$ is $1/4$ and the constant
is $3/4$, reflecting the fact that the formation of the Kondo singlet
influences only the contribution from the singlet channel. In Ref. \cite{12}%
, the same functional form as Eq.(7) is used for $G(T)$, however, the
corresponding prefactor and constant are both $1/2$. But the range in which
the measured points can be fitted well by a universal function is only from $%
1/2$ to $1$, as can be seen from Fig. 2(b) in Ref. \cite{12}. This implies
that the experimental data only confirm the $1/4$ of the conductance that is
influenced by the Kondo effect. The other $1/4$ may also vary with
temperature but perhaps due to other physical influences. In Ref. \cite{17},
from the data of quantum dots the prefactor is $1$ and the constant is $0$,
implying the full Kondo effect for the tunneling. This may indicate the
consequence of different structures between a QPC and a quantum dot: the
Kondo impurity is an extra local level in QPC as studied in this work while
it may be embedded in the tunneling path in quantum dots. It is known that
there is no anti-resonance associated with the latter case \cite{XRWang}. It
is easy to see that at low $T$ $F(\widetilde{T})\sim
1-p_{k}(T)=1-tanh(-T/T_{K})^{2}$. In the lower inset of Fig. 3 we show
(solid line) the scaling conductance of Eq.(\ref{scal}). For a comparison,
the experimentally obtained scaling conductance (dashed line) is also shown.
The experimental conductance is obtained from Eq.(1) and Eq.(2) of Ref. \cite%
{12}. The experimental $T_{K}$ is twice the value of the theoretical one.

By applying a bias voltage $V_b$ between two leads the current is $I=\frac{2e%
}{h}\int d\epsilon [f(\epsilon-eV_b/2)-f(\epsilon+eV_b/2)]\tau (\epsilon)$,
function $F(\widetilde{T})$ in Eq.(\ref{scal}) becomes $F(\widetilde{T}) \exp%
[-(eV_b/w)^2]$, reflecting the width $w$ ($\sim k_BT_K$) of the Kondo peak
in the spectral density \cite{Meir2}. We calculate the differential
conductance $g=\frac{dI}{dV_b}$, shown in the upper inset of Fig. 3. The
zero-bias peak, originated from the Fermi-level dependence of the Kondo
effect, exists only at low temperatures for which the 0.7 shoulder is
complemented to 1. By applying a magnetic field, the degeneracy of spins is
lifted, and the spin of the local level is along the field direction. As a
result, the singlet-triplet representation is no longer suitable and all the
plateaus become multiples of $0.5 (2e^2/h)$ due to the splitting of spin
subbands.

In summary, we demonstrate that a combination of anti-resonance of the
singlet channel and Kondo physics provides a satisfying account of basic
features associated with the "0.7 anomaly" in quantum point contacts.

We thank Junren Shi for helpful conversations. This work is supported by DOE
Grant No. DE/FG02-04ER46124, NSF-EPSCoR, and NSF-China.


\end{document}